%% file: fundamental.tex
\documentclass[a4paper, 10pt]{article}

\newcommand{\pdftitle}{The fundamental representation of pricing adjustments}
\newcommand{\pdfauthors}{Benedict Burnett, Ryan McCrickerd \& Benjamin Piau}
\newcommand{\pdfkeywords}{Derivatives, model risk, pricing, P\&L, volatility, XVA}

\title{
	\pdftitle\thanks{Disclaimer: This article represents the views of the authors alone and not the views of any firm or institution.}
	\vspace{3mm}
}
\author{
	Benedict Burnett\thanks{ben.burnett@barclays.com} \and
	Ryan McCrickerd\thanks{ryan.mccrickerd@barclays.com} \and
	Benjamin Piau\thanks{benjamin.piau@barclays.com}
	\vspace{3mm}
}
\date{
	Barclays QA XVA \\
	\vspace{7mm}
	22nd July, 2026
}

\usepackage{tikz}
\usepackage[nodayofweek, level]{datetime}
\usepackage{amsmath}
\usepackage{amssymb}
\usepackage{amsthm}
\usepackage{mathtools}
\numberwithin{equation}{section}

\usepackage[pdftex,
	bookmarks=true,
	unicode=true,
	pdftoolbar=false,
	pdfmenubar=false, 
	pdffitwindow=true,     
	pdfstartview={FitV},
	pdftitle={\pdftitle},   
	pdfauthor={\pdfauthors},    
	pdfsubject={},  
	pdfcreator={},   
	pdfproducer={}, 
	pdfkeywords={\pdfkeywords}, 
	pdfnewwindow=true,     
	colorlinks=true,      
	linkcolor=black,         
	citecolor=black,
	filecolor=black,
	urlcolor=black
]{hyperref}

\usepackage[longnamesfirst]{natbib}
\usepackage[justification=centering,font=small]{caption}
\usepackage[multiple]{footmisc}

\newcommand{\bb}[1]{{\mathbb{#1}}}
\newcommand{\cc}[1]{{\mathcal{#1}}}
\newcommand{\rr}[1]{{\mathrm{#1}}}

\newcommand{\qq}{\bb{Q}}
\newcommand{\hp}{{\hat{\bb{P}}}}
\newcommand{\hq}{{\hat{\bb{Q}}}}
\newcommand{\hV}{{\hat V}}
\renewcommand{\ll}{\cc{L}}
\newcommand{\hl}{\hat{\cc{L}}}
\newcommand{\bleed}{{Z}}

\newcommand{\pd}{\partial}

\begin{document}

\maketitle

\begin{abstract}
	This article consolidates and extends past work on derivative pricing adjustments, including XVA, by providing an encapsulating representation of the adjustment between any two derivative pricing functions, within an It\^o SDE/parabolic PDE framework.
	We give examples of this representation encapsulating others from the past 20 years, ranging from a well known option pricing adjustment introduced by Gatheral, to the collection of semi-replication XVA originating from Burgard \& Kjaer.
	We show that this fundamental representation can be applied to quantify and mitigate XVA model risk, providing a novel approach to estimating CVA wrong-way risk as an example application.
\end{abstract}

\vspace{5mm}
\hspace{3.5mm}Keywords: \pdfkeywords

\vspace{1mm}
\hspace{3.5mm}MSC2020: 91G20, 91G30, 91G40

\clearpage
\setcounter{tocdepth}{2}
\tableofcontents

\clearpage
\section{Introduction}\label{sec:intro}

There has been a large amount of work in the literature on adjustments to `naive' derivative prices, ranging from early work presented in e.g.~\cite{gatheral_2006}, through \cite{piterbarg_2010}, \cite{burgard_2013} and \cite{green_2014} to, more recently, work such as \cite{burnett_2021a} and \cite{kenyon_2023_co2}. These articles have demonstrated the possibilities that are opened by accommodating sophisticated effects (be it volatilities, funding, hedging or climate costs) via adjustments to well-understood base calculations. What has perhaps not been explicitly addressed is the relationship between all of these adjustments, placing them within a unified view of the landscape. That is the primary aim of the present article. We will see that valuation adjustments can generally be described in terms of three types of contribution: model adjustments, discounting adjustments, and payoff adjustments. Figure \ref{fig:venn} shows this conceptual map, highlighting the locations of some key cases from the literature.

\begin{figure}[htbp]
	\centering
	\input{picture}
	\caption{The pricing adjustments universe as suggested by \eqref{eq:intro-adjustment}. Citations show where historic examples live, analysed in Section \ref{sec:examples}. The location of \cite{piterbarg_2010} here refers strictly to the fully-collateralised case.}
	\label{fig:venn}
\end{figure}

This theoretical unification has a practical purpose, which leads to the secondary aim of this article: it suggests a new and general way to estimate the impact of model changes (such as adding a new stochastic factor, or changing the assumed dynamics in another way). This approach can be applied to cases including XVAs, where such impacts can be particularly relevant (see \cite{benezet_2024}). Since XVAs are already adjustments to underlying valuations, in the below, we term any adjustments on XVAs themselves `meta-adjustments'.

The results of this investigation have multiple potential beneficiaries. The formulae we give for general adjustments show a way to estimate future P\&L bleed (important for traders, finance functions and senior management); equivalently, they suggest ways to approximate the inaccuracy in a simple calculation due to neglected effects (important for model validation and valuation control).

\subsection{Structure of the article}
In Section \ref{sec:fundamental} we present a fundamental representation of pricing adjustments. This representation reveals a natural decomposition of the pricing adjustments universe, in terms of three main categories (M, D and P) relating to models, discounting, and payoffs.\footnote{%
	This set of effects is natural given the ingredients in basic pricing rules, e.g.~$\bb{E}[e^{-rT}G(X_T)]$. Notice the \emph{model} $X_t$, \emph{discounting} at rate $r$, and \emph{payoff} $G$, which all stand to be interrogated.%
}\footnote{%
We thank an anonymous reviewer for the suggestion that an alternative view would combine D and P effects into `cashflows', in contrast to dynamics (M); it is possible that this approach might yield a more frequently distinct separation of adjustments than the granular approach we take here.} 

Considering a desirable target price $\hat V$ in terms of a convenient base price $V$ plus some adjustment $U$, this categorisation derives from the associated P\&L bleed $\bleed$ on which $U$ depends. Summarising what we will show in Section \ref{sec:fundamental},
\begin{equation}\label{eq:intro-adjustment}
	U=\bb{E}\left[\int_0^Te^{-\int_0^t\hat R_s ds}\bleed_t \,dt\right],\quad \bleed = \underbrace{(\hl - \ll)}_{\neq0\Leftrightarrow\rr{\textbf{M}}}V - \underbrace{(\hat R - R)}_{\neq0\Leftrightarrow\rr{\textbf{D}}}V + \underbrace{(\hat F - F)}_{\neq0\Leftrightarrow\rr{\textbf{P}}} .
\end{equation}
As described in Table \ref{tab:notation}, $\ll$, $R$ and $F$ are the (naive) It\^o generator of a process $X_t$, discount rate, and payoff respectively, all reflected in the base price $V$. Terms with hats are counterparts from $\hat V$. 

Although in principle the adjustments covered here might span multiple teams in financial institutions,\footnote{A simple model (e.g.~convexity) adjustment could be handled directly by an origination desk, a payoff adjustment (e.g.~CVA) could be handled by the XVA desk, etc.} for concreteness we will henceforth imagine that a bank sells a derivative to a client at the target price $\hat V$, then distributes its risk management between an origination desk handling $V$ and XVA desk handling $U$. But this should be taken with a pinch of salt, i.e.~idealised abstraction.

Moving to XVA model risk in Section \ref{sec:meta}, $U$ plays the role of $V$ and we call $A=\hat U - U$ a meta-adjustment.\footnote{%
	Although we explicitly consider only one level of meta-adjustment in this article, the approach can in principle be taken to higher orders, with adjustments on adjustments on adjustments (and so on) without needing any amendments to the theory presented here. We thank an anonymous reviewer for pointing this out.%
} Due to the extreme difficulty of computing these accurately in many cases, the theory developed here serves as a robust framework for constructing signals for adjustments' magnitude and direction.\footnote{In 2001 Gatheral wrote \emph{it is no surprise that this approach doesn't yield any easy results}, essentially regarding $U$ in \eqref{eq:example-intro}. \emph{Easy} would certainly be an understatement for meta-adjustments, depending on future XVA greeks, but this does not preclude valuable results.} Based around expected P\&L, these can be vital towards banks understanding risk \emph{today}, given model upgrades can take significant time. We show that the general representation derived in Section \ref{sec:fundamental} provides a surprisingly effective tool to approximate hard-to-calculate meta-adjustments, demonstrating its power on the classic case of estimating wrong-way risk on a CVA.

\subsection{A starting example}
The region M of `model adjustments' in Figure \ref{fig:venn} is frequently the most challenging, being driven by changes to the model dynamics, so it serves to consider an example from \cite{gatheral_2006}. In his equation (3.4), Gatheral writes a call option price $\hat V$ on a stock $S$ in terms of a Black-Scholes price $V$ plus some adjustment $U$:
\begin{equation}\label{eq:example-intro}
	\hV = V + U,\quad U=\bb{E}\left[\int_0^T \!\bleed_t \,dt\right],\quad \bleed=\frac12\left(\hat \alpha^2 - \alpha^2\right)S^2 \partial_{SS}V.
\end{equation}
The variable $\int_0^T \bleed_t \,dt$ here represents the (pathwise) P\&L after selling (and delta hedging) the option using the Black-Scholes model with volatility $\alpha$, while $S$ \emph{actually} evolves with volatility $\hat \alpha$. Its representation in terms of the option gamma $\partial_{SS}V$ is often referred to as the \emph{fundamental theorem of derivative trading}.\footnote{%
	See \cite{ellersgaard_2017} for the history going back to \cite{karoui_1998}, with the experiments of \cite{ahmad_2005} deserving special mention. The associated \emph{pricing adjustment} $U$ was, to the best of our knowledge, first presented in \cite{gatheral_2001}, then refined in \cite{lee_2005}.%
} (Comparing with \eqref{eq:intro-adjustment}, the presence of only one term in formula \eqref{eq:example-intro} for $\bleed$ stems from the fact that this case covers a pure model adjustment, with no discount or payoff contributions.)

In this example, the adjustment specifically reflects Black-Scholes model risk for a call option. What we present in this article both extends this model adjustment example to its logical conclusion within an It\^o SDE/parabolic PDE framework, and unifies it with the other discounting and payoff adjustments that are typical in XVA.\footnote{%
	While this extension/unification is clearly quite general, it does not cover \emph{all} models and derivative types. American-style options are excluded (at least without approximations) since they do not solve parabolic PDEs on their entire domain.%
} To accompany \cite{gatheral_2006}, for these others we use familiar examples from \cite{piterbarg_2010} and \cite{burgard_2013}.

\subsection*{Notation}\label{sec:notation}

Table \ref{tab:notation} summarises our notation. Everything up to the base price $V$ is associated with a probability measure $\qq$, and has a counterpart $\hat\sigma$, $\hl$ etc.~relating to $\hat V$ and $\hq$. All \emph{functions} $f=f(t,x)$ defined on $\cc{X}$ induce \emph{processes} $f_t:=f(t,X_t)$.

\begin{table}[htbp]
	\begin{center}
		\begin{tabular}{ll}
			Symbol & Description \\ \hline\rule{0mm}{4mm}\!\!
			$X_t$ & $n$-dim It\^o process, with $d X_t=\mu_t dt + \sigma_t dW_t$ \\
			$\mu=\mu(t,x)$ & $n$-dim drift vector \\
			$\sigma=\sigma(t,x)$ & $n\times d$ diffusion matrix \\
			$W_t$  & $d$-dim Brownian motion \\
			$\rho$ & $d\times d$ correlation matrix, with $d\langle W \rangle_t=\rho \,dt$ \\
			$a=a(t, x)$ & $n\times n$ covariation matrix, with $d\langle X \rangle_t=a_t dt$ \\
			$\ll$ & It\^o generator of $X_t$, with $\ll=\mu\cdot\partial_x + \frac12 a:\partial_{xx}$ \\
			$\partial_x,\partial_{xx}$ & Gradient, Hessian, e.g.~$(\partial_{xx})_{ij}=\partial_{ij}=\frac{\partial^2}{\partial x_i\partial x_j}$ \\
			$a:b$ & Double-dot product, i.e.~$\sum_{ij}a_{ij}b_{ij}=\rr{tr}\left(ab^\rr{T}\right)$\\ 
			$R=R(t,x)$ & Discounting function \\		
			$F=F(t,x)$ & Continuous payoff function \\		
			$G=G(x)$ & Terminal payoff function \\
			$V=V(t,x)$ & Pricing function \\
			$\cc{X}$ & Pricing function domain $[0,T]\times \bb{R}^n$ \\
			$U=U(t,x)$ & Pricing adjustment $\hat V - V$ \\
			$A=A(t,x)$ & Meta-adjustment $\hat U - U$ \\
			$\bleed=\bleed(t,x)$ & P\&L bleed \\
			$\bb{E}_t[\cdot]$, $\bb{E}_{t,x}[\cdot]$ & Conditional expectations $\bb{E}[\cdot|X_t]$, $\bb{E}[\cdot|X_t=x]$
		\end{tabular}
		\caption{Main notation.}
		\label{tab:notation}
	\end{center}
\end{table}

\section{Fundamental representation}\label{sec:fundamental}

We seek a convenient representation $V + U$ of a target price $\hat V$. By \emph{convenient} we essentially mean that the adjustment $U$ should depend only on $V$ directly, not on $\hat V$. Both $V$ and $\hat V$ are assumed to be suitably well-defined on a common domain $\cc{X}:=[0,T]\times\bb{R}^n$.\footnote{%
	An alternative approach would be to consider the domain of $\ll$ explicitly to be a subset of that of $\hl$. However, we can take them both to be defined on the common superset $\cc{X}$ essentially w.l.o.g., given we can e.g.~consider $V$ to depend trivially on variables. %
} The adjustment $U=\hat V-V$ is then well-defined too, hence our particular focus on its \emph{representation}.

Specifically, we assume that $V$ satisfies the following parabolic PDE
\begin{equation}\label{eq:base-pde}
	\partial_t V + \cc{L} V - RV + F=0,\quad V(T,\cdot) = G,
\end{equation}
where the operator $\ll$ is given by 
\begin{equation*}
	\cc{L} := \sum_{i=1}^n \mu_i\partial_i + \frac12\sum_{i=1}^n\sum_{j=1}^n a_{ij}\partial_{ij},\quad a_{ij} := \sum_{k=1}^d\sum_{l=1}^d \rho_{kl}\sigma_{ik}\sigma_{jl}.
\end{equation*}
This is the It\^o generator of a process $X_t$ under a measure $\qq$, which we will get to shortly. We will write $\ll=\mu\cdot\partial_x + \frac12 a:\partial_{xx}$ for this going forward.\footnote{Note that in all examples presented, one-factor models will be used, meaning $d=n$ and $\sigma_{ij}=0$ for $i\neq j$. Then we have the simplification $a_{ij}=\rho_{ij}\sigma_i\sigma_j$, where $\sigma_i:=\sigma_{ii}$.} As summarised in Table \ref{tab:notation}, these functions $\mu$, $\sigma$, $F$ etc.~are all defined on $\cc{X}$ like $V$. 

Now we assume that the target price $\hat V$ instead solves a different PDE
\begin{equation}\label{eq:target-pde}
	\partial_t \hat V + \hl \hat V - \hat R\hat V + \hat F=0,\quad \hat V(T,\cdot) = \hat G,\quad \hl=\hat\mu\cdot\partial_x + \frac12 \hat a:\partial_{xx}.
\end{equation}
Notice that within the class of all such PDE solutions, the similarities of $V$ and $\hat V$ end with their common domain $\cc{X}$, i.e.~they are \emph{very} loosely related.

By subtracting \eqref{eq:base-pde} from \eqref{eq:target-pde} and eliminating $\hat V$, we see that $U=\hat V-V$ solves the PDE
\begin{equation*}
	\partial_t U + \hl U + (\hl - \ll)V - \hat R U - (\hat R - R)V + \hat F - F = 0,\quad U(T,\cdot) = \hat G - G.
\end{equation*}
We can write this more succinctly as
\begin{equation}\label{eq:adjustment-pde}
	\partial_t U + \hl U - \hat RU + F^*=0,\quad U(T,\cdot) = G^*,
\end{equation}
where
\begin{equation}\label{eq:adj-source-payoff}
	F^* :=  (\hl - \ll) V -(\hat R - R)V + \hat F - F,\quad G^* := \hat G - G
\end{equation}
and
\begin{equation*}
	(\hl - \ll) := (\hat\mu - \mu)\cdot\partial_x + \frac12(\hat a - a):\partial_{xx}.
\end{equation*}
By applying the Feynman-Kac theorem to \eqref{eq:adjustment-pde},\footnote{See e.g.~\cite{karatzas_1998} Theorem 7.6 for some conditions to legitimise this.} we obtain the \emph{fundamental representation} of the adjustment $U$, namely
\setlength{\fboxsep}{2mm}
\begin{equation}\label{eq:fundamental-representation}
	\boxed{
		U(t,x) = \bb{E}_{t,x}^\hq\left[\int_t^T e^{-\int_t^u \hat R(s, X_s)ds}F^*(u, X_u)\,du + e^{-\int_t^T \hat R(u, X_u)du} G^*(X_T)\right].
	}
\end{equation}

This stochastic representation for $U$ is the unique one not depending on $\hat V$ modulo trivialities, i.e.~to the extent that the PDE in \eqref{eq:adjustment-pde} is likewise unique.\footnote{If we accept dependence on $\hat V$ then we can e.g.~choose more convenient discounting $R^*$ in \eqref{eq:adjustment-pde}, at the expense of more complicated $F^*$, as in \cite{elouerkhaoui_2016}.} The price we pay for this is that the process $X_t$ is evaluated under the target measure $\hq$ that is related to $\hat V$ and not $V$. Specifically, $\hq$ is a measure under which the $n$-dimensional It\^o process $X_t$ has generator $\hl$, i.e.~dynamics
\begin{equation*}
	d X_t = \hat \mu(t, X_t)dt + \hat \sigma(t, X_t)d W_t,\quad d\langle W^i,W^j\rangle_t = \hat \rho_{ij}dt,
\end{equation*}
in terms of a $d$-dimensional Brownian motion $W_t$ with correlation matrix $\hat \rho$.\footnote{See \cite{karatzas_1998} (7.2)-(7.4) for well-posedness conditions of such It\^o SDEs.} Importantly, note that there is no requirement that $\qq$ and $\hq$ be \emph{equivalent} measures (in particular the volatility processes $\sigma_t$ and $\hat \sigma_t$ may be quite different).

We will write this representation in \eqref{eq:fundamental-representation} more succinctly as
\begin{equation}\label{eq:succinct-repr}
	U = \bb{E}_{t,x}^\hq\left[\int_t^T e^{-\int_t^u \hat R_sds}F^*_u \,du + e^{-\int_t^T \hat R_udu} G^*_T\right],
\end{equation}
and similarly $d X_t = \hat \mu_t dt + \hat \sigma_t d W_t$, $d\langle W\rangle_t = \hat\rho \,dt$.
Moreover, since the terminal payoff $G^*$ ends up teaching us little more than can be inferred from $F^*$,\footnote{E.g.~we can reinstate $G^*$ later by letting $F^*_tdt\mapsto F^*_tdt + G^*_t\delta_T(dt)$ for Dirac measure $\delta_T$.} we henceforth assume $\hat G= G$ and thus $G^*=0$. As covered in Appendix \ref{app:pnl}, the function $F^*$ then fully captures the P\&L bleed $\bleed$ from pricing and hedging $V$ in a world that is instead consistent with $\hat V$, i.e.~which evolves under some $\hp\sim\hq$. 

So as summarised in Section \ref{sec:intro}, and depicted in Figure \ref{fig:venn}, we obtain
\begin{equation}\label{eq:fundamental}
	U=\bb{E}_{t,x}^\hq\left[\int_t^Te^{-\int_t^u\hat R_s ds}\bleed_u \,du\right],\quad \bleed := (\hl - \ll)V - (\hat R - R)V + (\hat F - F).
\end{equation}

Note that while we defined \emph{model} adjustments in M to mean $\hl\neq\ll$ as in \eqref{eq:example-intro}, there is nothing to stop e.g.~\emph{payoff} adjustments with $\hat F\neq F$ arising purely from model \emph{dependence}. We elaborate on this in Appendix \ref{app:model-dependence}.

Taking stock, our adjustment $U$ takes us between two very general pricing functions $V$ and $\hat V$. These clearly also admit stochastic representations like $U$ in \eqref{eq:succinct-repr}, which in practice might define them. E.g.~$V$ may be written 
\begin{equation}\label{eq:base-price-sde}
	V = \bb{E}_{t,x}^\qq\left[\int_t^T e^{-\int_t^u R_s ds}F_u du + e^{-\int_t^T R_u du} G_T  \right]
\end{equation}
where $\qq\not\sim\hq$ is a measure where $X_t$ instead has It\^o generator $\ll$, i.e.~dynamics
$$d X_t = \mu_t dt + \sigma_t d W_t,\quad d\langle W\rangle_t = \rho dt.$$

So even the base price $V$ is general enough to cover exotic derivatives, driven by non-tradable risk factors. This extends the frameworks of \cite{bossy_2006} and \cite{ellersgaard_2017}, and covers meta-adjustments as in Section \ref{sec:meta}.

\subsection*{Summary}

When giving specific adjustment examples, we need to state \emph{either} the PDE for the base price $V$ as in \eqref{eq:base-pde}, \emph{or} (equivalently) the stochastic representation as below, and likewise for the target price $\hat V$. Then \eqref{eq:succinct-repr} gives the adjustment $U$. In summary, using stochastic price representations,
\begin{align*}
	V &= \bb{E}_{t,x}^\qq\left[\int_t^T e^{-\int_t^u R_s ds}F_u \,du + e^{-\int_t^T R_u du} G_T  \right],\\
	\hat V &= \bb{E}_{t,x}^\hq\left[\int_t^T e^{-\int_t^u \hat R_s ds}\hat F_u \,du + e^{-\int_t^T \hat R_u du} \hat G_T  \right],\\
	&\implies \hat V - V = U = \bb{E}_{t,x}^\hq\left[\int_t^T e^{-\int_t^u \hat R_sds}F^*_u\,du + e^{-\int_t^T \hat R_udu} G^*_T\right],
\end{align*}
where $F^*$ and $G^*$ are defined in \eqref{eq:adj-source-payoff}. For model adjustments ($\hl\neq\ll$) we can summarise the dynamics of $X_t$ under $\qq$ and $\hq$ by
$$
dX_t=\mu_t \,dt + \sigma_t dW_t,\ d\langle W\rangle_t=\rho \,dt
\quad \xmapsto{\qq\mapsto\hq}\quad
dX_t=\hat \mu_t \,dt + \hat\sigma_t dW_t,\ d\langle W\rangle_t=\hat \rho \,dt.
$$
We reiterate the fact that there is no assumption that $\qq$ and $\hq$ be equivalent -- the most challenging adjustments to compute arise precisely from cases where they are not (e.g.\ where one measure introduces additional risk factors).

\section{Historic adjustment examples}\label{sec:examples}

In this section we show how our fundamental representation of adjustments from \eqref{eq:fundamental-representation} encapsulates a range of others from the past. With the P\&L bleed breakdown from \eqref{eq:intro-adjustment} in mind, our main focus is on the \emph{model}, \emph{discounting} and \emph{payoff} adjustment examples shown in Figure \ref{fig:venn}. Since we will have $G^*=0$, all adjustments here will reduce to the form of \eqref{eq:fundamental}.

\subsection{Model: Gatheral (2006)}\label{sec:gatheral}

This example is from Chapter 3 of \cite{gatheral_2006}, and was discussed briefly in Section \ref{sec:intro}. 
This is a case of a strict model adjustment from Figure \ref{fig:venn}, specifically $\hl\neq\ll$, $\hat R=R=0$, $\hat F=F=0$ and common European payoff $\hat G=G$. In summary therefore,
\begin{equation*}
	V=\bb{E}_{t,x}^\qq\left[G_T\right], \quad \hat V=\bb{E}_{t,x}^\hq\left[G_T\right],\quad U=\bb{E}_{t,x}^\hq\left[\int_t^T\bleed_u \,du\right],\quad \bleed = (\hl - \ll)V.
\end{equation*}

All we have to do here is specialise the general operator's representation,
\begin{equation*}
	(\hl-\ll)=(\hat\mu-\mu)\cdot\partial_x + \frac12(\hat a - a):\partial_{xx},
\end{equation*}
which is eased by the use of one-factors models here, since then e.g.~$a_{ij}=\rho_{ij}\sigma_{ii}\sigma_{jj}$. We cover local volatility ($n=1$) and stochastic volatility ($n=2$) separately, since our representation provides a different and noteworthy perspective on the latter.

\subsubsection*{Local volatility}

For local volatility, we fix dimensions $d=n=1$, denote a stock price variable $x=S$, process $X_t=S_t$ and functions $\sigma=\sigma(t,S)$, $V=V(t,S)$ etc. Encode driftless Black-Scholes and local volatility dynamics of $S_t$ under $\qq$ and $\hq$ respectively through the functions
\begin{equation*}
	\mu = 0,\quad
	\sigma = \alpha S;\quad
	\hat\mu = 0,\quad
	\hat\sigma= \hat\alpha S,
\end{equation*}
for \emph{constant} Black-Scholes volatility $\alpha>0$, and local volatility \emph{function} $\hat\alpha=\hat\alpha(t,S)$. Now the adjustment $U=\hat V-V$ purely reflects volatility risk, i.e.~from
$$
dS_t=\alpha S_t dW_t\quad \xmapsto{\qq\mapsto\hq}\quad dS_t=\hat\alpha_t S_t dW_t.
$$
This yields P\&L bleed $Z = \frac12 (\hat\alpha^2 - \alpha^2)S^2\partial_{SS}V$, so e.g.~$U_0=\hat V_0-V_0$ is given directly by
\begin{equation}\label{eq:local-vol-adj}
	U_0 = \bb{E}_0^\hq\left[\int_0^T \frac12(\hat\alpha_t^2 - \alpha^2)S_t^2(\partial_{SS}V)_t \,dt\right].
\end{equation}

\subsubsection*{Stochastic volatility}

As in \cite{bergomi_2016} Section 2.4.1, the above generalises to accommodate going from local to stochastic volatility with \emph{any} dimension $n> 1$, and again we get $U_0$ as in \eqref{eq:local-vol-adj}.
However then $\hat\alpha$ becomes a \emph{variable} of $\hat V$ in its own right, unlike the \emph{parameter} $\alpha$ of $V$.\footnote{This inconsistent volatility treatment is not in the spirit of this work, since we \emph{explicitly} consider $V$ and $\hat V$ on a common domain $\cc{X}$. Treating $\alpha$ as a variable under $\qq$ instead, with associated `process' $d\alpha_t=0$, has tangible consequences. (Think about the associated hedging strategy in Appendix \ref{app:pnl}.) This will come up again in Section \ref{sec:meta}, regarding XVA parameter risk.} Fixing $n=2$ for simplicity, instead let $x=(S,\alpha)$ and treat $\alpha$ as a variable of both $V$ and $\hat V$, with associated dynamics 
$$
dS_t=\alpha_t S_t dW^S_t,\quad d\alpha_t=0\quad \xmapsto{\qq\mapsto\hq}\quad dS_t=\alpha_t S_t dW^S_t,\quad d\alpha_t=\hat\beta_tdt+\hat\gamma_tdW^\alpha_t,
$$
and common correlation $\hat\rho_{S\alpha}=\rho_{S\alpha}$.\footnote{E.g.~for Heston dynamics choose $\hat\beta$ and $\hat\gamma$ such that $d\alpha^2_t=\kappa(\theta-\alpha^2_t)dt+v\alpha_tdW^\alpha_t$.} Now we explicitly have 2d functions
\begin{equation*}
	\mu =
	\begin{pmatrix}
		0\\
		0
	\end{pmatrix},\quad
	\sigma =
	\begin{pmatrix}
		\alpha S & 0\\
		0 & 0
	\end{pmatrix},\quad
	\hat\mu =
	\begin{pmatrix}
		0\\
		\hat\beta
	\end{pmatrix},\quad
	\hat\sigma=
	\begin{pmatrix}
		\alpha S & 0\\
		0 & \hat\gamma
	\end{pmatrix},
\end{equation*}
and obtain P\&L bleed $Z = \hat\beta\,\partial_\alpha V + \frac12 \hat\gamma^2\,\partial_{\alpha\alpha}V + \rho_{S\alpha}\alpha S\hat\gamma\,\partial_{S\alpha}V$, thus adjustment
\begin{equation*}
	U_0 = \bb{E}_0^\hq\left[\int_0^T \left(\hat\beta_t(\partial_\alpha V)_t + \frac12 \hat\gamma_t^2(\partial_{\alpha\alpha}V)_t + \rho_{S\alpha}\alpha_t S_t\hat\gamma_t(\partial_{S\alpha}V)_t\right) dt\right].
\end{equation*}
Now the gamma term $\partial_{SS}V$ cancels since its coefficient $\hat\sigma^2_{11}-\sigma^2_{11}$ does, and here we have Black-Scholes vega $\partial_\alpha V$, vanna $\partial_{S\alpha}V$ and volga $\partial_{\alpha\alpha}V$ terms instead. Implicit in this P\&L bleed $Z$ is a better hedging strategy, compared with \eqref{eq:local-vol-adj}, so we expect a better variance reduction compared with \cite{bergomi_2016} Appendix 8.A.2. (This is indeed seen in simple numerical experiments.)

\subsection{Discounting: Piterbarg (2010)}

Here we consider strict discounting adjustments, starting with $\hl=\ll$, $\hat R\neq R$, $\hat F=F=0$ and $\hat G=G\neq0$ with deterministic IR basis as in \cite{piterbarg_2010}. Then we have prices,
\begin{equation*}
	V=\bb{E}_{t,x}^\qq\left[e^{-\int_t^T R_u du}G_T\right], \quad \hat V=\bb{E}_{t,x}^\qq\left[e^{-\int_t^T \hat R_u du}G_T\right],
\end{equation*}
and we have P\&L bleed $\bleed = -(\hat R - R)V$, so the adjustment $U_0 = \hat V_0 - V_0$ is given by
\begin{equation}\label{eq:disc-adj}
	U_0=-\bb{E}_0^\qq\left[\int_0^Te^{-\int_0^t \hat R_s ds}(\hat R_t - R_t)V_t \,dt\right].
\end{equation}
This expression coincides with \cite{piterbarg_2010} Equation (3),\footnote{Specifically for the fully collateralised case $C=V$, with notation $\hat R=r_F$, $R=r_C$. The partially collateralised case is best accommodated as a payoff adjustment (i.e.~a COLVA).} but the scope here is wider given our unconstrained dimension and dynamics of $X_t$, driving everything e.g.~rates $\hat R_t=\hat R(t,X_t)$.\footnote{Indeed, using $\hat R_t=\max(r^1_t, \dots,r^m_t)$ we cover CTD CSAs as in \cite{kemarsky_2025}.} Notice that the form of \eqref{eq:disc-adj} is actually unchanged if we allow exotic $V$ and $\hat V$, provided that $\hat R\neq R$ only, e.g.
\begin{multline*}
	V = \bb{E}_{t,x}^\qq\left[\int_t^T e^{-\int_t^u R_s ds}F_u \,du + e^{-\int_t^T R_u du} G_T  \right],\\
	\hat V = \bb{E}_{t,x}^\qq\left[\int_t^T e^{-\int_t^u \hat R_s ds} F_u \,du + e^{-\int_t^T \hat R_u du} G_T  \right].
\end{multline*}
This also generalises to the funding invariance principle of \cite{elouerkhaoui_2016}.

\subsection{Payoff: Burgard \& Kjaer (2013)}\label{sec:burgard}

Here we will recover the CVA formula presented in \cite{burgard_2013} Section 3.3 (`Strategy II'), which as in Figure \ref{fig:venn} lives in the intersection of discounting and payoff adjustments, specifically with $\hl=\ll$, $\hat R\neq R$, $\hat F\neq F=0$ and $\hat G=G\neq0$. Such XVA examples are usually the simplest mathematically.\footnote{Recall \cite{elouerkhaoui_2016}: it is usually just a matter of specifying the exact \emph{payoff}.} 

Contrasting the previous examples, for XVA we typically define $V$ and $\hat V$ through PDEs. So specialising the PDEs in \eqref{eq:base-pde} and \eqref{eq:target-pde}, assume
\begin{equation*}
	\partial_t V + \cc{L} V - RV = 0,\quad \partial_t \hat V + \ll \hat V - \hat R\hat V + \hat F=0,\quad \hat V(T,\cdot) = V(T,\cdot) = G,
\end{equation*}
then the PDE in \eqref{eq:adjustment-pde} for the adjustment $U=\hat V-V$ reduces to 
\begin{equation*}
	\partial_t U + \ll U - \hat RU + F^* = 0,\quad U(T,\cdot) = 0,
\end{equation*}
and from \eqref{eq:succinct-repr} we get the adjustment's stochastic representation,
\begin{equation*}
	U = \bb{E}_{t,x}^\qq\left[\int_t^T e^{-\int_t^u \hat R_sds} F^*_u\,du \right].
\end{equation*}
Specialising this further, introduce an interest rate and counterparty hazard rate $x=(r, \lambda, x_3, \dots, x_n)$, assume $R=r$, and that we have discounting difference $\hat R - R = \lambda$ and source term $F^* = -(\hat R - R) V + (\hat F - F) = -\lambda V^+$.\footnote{These differences arise from hedging a loss of $V^+:=\max(V,0)$ upon counterparty default.} Then we obtain
\begin{equation*}
	\partial_t U + \ll U - (r + \lambda)U -\lambda V^+ = 0,\quad U = -\bb{E}_{t,x}^\qq\left[\int_t^T e^{-\int_t^u (r_s + \lambda_s) ds}\lambda_u V^+_udu \right],
\end{equation*}
which coincides with \cite{burgard_2013} Equations (36) and (37).\footnote{We have assumed zero recovery, funding spread and collateral, for simplicity.}

Similarly we can reconcile \emph{strict} payoff adjustments such as KVA from \cite{green_2014} or HVA from \cite{burnett_2021a}. E.g.~write $\hat F=F+f$ for capital or friction cost $f$, with all else equal, to obtain $U = \bb{E}_{t,x}^\qq[\int_t^T e^{-\int_t^u R_sds} f_udu ]$.

\section{Meta-adjustments}\label{sec:meta}

\subsection{The concept}
We now focus on XVA model risk. To this end, let $U$ represent some base XVA price, then let $A$ denote a meta-adjustment that takes us to some target XVA price $\hat U = U + A$. XVAs are of course already adjustments (to some price $V$), conventionally residing in the sets of discounting ($\hat R\neq R$) and/or payoff adjustments ($\hat F\neq F$) in Figure \ref{fig:venn}, but \emph{not} model adjustments ($\hl = \ll$). Focusing on novelties, we consider here some \emph{strict} model adjustments to XVA, i.e.~with $\hat R=R$, $\hat F=F$ and $\hl\neq\ll$. So $A$ lives alongside \cite{gatheral_2006} in Figure \ref{fig:venn}; $U$ lives with \cite{burgard_2013}, while $\hat U$ lives in the very centre.\footnote{%
	Compare here the CO2eVA of \cite{kenyon_2023_co2}, which can be seen to be a payoff meta-adjustment driven by the $G^*$ terminal payoff term from Section \ref{sec:fundamental}.
}

Keeping our CVA example from Section \ref{sec:burgard} in mind, we suppose that
\begin{equation}\label{eq:meta-prices}
	U=\bb{E}_{t,x}^\qq\left[\int_t^Te^{-\int_t^u R_s ds}F_u \,du\right],\quad \hat U=\bb{E}_{t,x}^\hq\left[\int_t^Te^{-\int_t^u R_s ds} F_u \,du\right],
\end{equation}
so that $U$ and $\hat U$ differ through the dynamics of $X_t$ only, implicit in the measure difference. Now adapting \eqref{eq:fundamental},\footnote{E.g.~by simply changing notation $(V,U)\mapsto(U,A)$. For convenience, we recycle notation $\sigma$, $F$, $\qq$ etc.~from $V$ to define $U$ here, since we will not need to focus on these details of $V$.} our fundamental representation of the \emph{meta}-adjustment $A=\hat U - U$ is
\begin{equation}\label{eq:fundamental-meta}
	A=\bb{E}_{t,x}^\hq\left[\int_t^Te^{-\int_t^u R_s ds}\bleed_u \,du\right],\quad \bleed := (\hl - \ll)U.
\end{equation}
Here $\bleed$ reflects our P\&L bleed function as usual, now specifically deriving from some XVA model misspecification, i.e.~the use of $X_t$ under $\qq$, with $\qq\not\sim\hq\sim\hp$ for real-world measure $\hp$. Since the compact \eqref{eq:fundamental-meta} is prone to disguising complexities, recall that in full, evaluated at $t=0$, this reads
\begin{equation}\label{eq:3}
	A_0=\bb{E}_0^\hq\left[\int_0^Te^{-\int_0^t R_s ds}\left( (\hat \mu - \mu)_t\cdot (\partial_x U)_t + \frac12 (\hat a - a)_t:(\partial_{xx}U)_t\right) dt\right].
\end{equation}

Depending on $O(n^2)$ future XVA greeks, computing such meta-adjustments accurately can present a serious challenge, and institutions will firstly be interested in indicative signals for this adjustment's magnitude and direction.\footnote{%
	Indeed, good XVA traders can often develop some such signals, e.g.~based on today's P\&L bleed $\bleed_0$ and the reasonable assumption that $\bleed_t\to0$ as $t\to T$ for XVA.%
} Relatedly, traders will be interested in estimating their expected P\&L if they were not to make and hedge such an adjustment, which is shown in Appendix \ref{app:pnl} to be given by \eqref{eq:fundamental-meta} but replacing $\hq$ by the real-world $\hp$.\footnote{Note that evaluating the P\&L bleed \emph{process} $\bleed_t=\bleed(t,X_t)$ under $\hp\sim\hq$ means $X_t$ has real drift $\hat\mu^*\neq\hat\mu$, while the \emph{function} $\bleed=\bleed(t,x)$ retains dependence on risk-neutral $\hat\mu$ via $\hl$.}

\subsection{P\&L bleed term-structure} 
Moving in this direction of estimations, we can first rewrite $A_0$ in the form of a deterministic integral, 
\begin{equation}\label{eq:pnl-bleed-ts}
	A_0 := \bb{E}^\hq_0\left[\int_0^T e^{-\int_0^t R_s ds}Z_t dt\right] = \int_0^T e^{-\int_0^t \hat R^*_0(s)ds}\hat Z^*_0(t)dt.
\end{equation}
Here $\hat R^*_0$ is a (generalised) forward rate term-structure, and $\hat Z^*_0$ a \emph{P\&L bleed term-structure}, respectively defined by
\begin{equation}\label{eq:pnl-bleed-defn}
	\hat R^*_0(t) := -\partial_t\log \bb{E}^\hq_0\left[e^{-\int_0^t R_sds}\right],\quad \hat Z^*_0(t) := \bb{E}_0^{\hq_t}\left[Z_t\right],
\end{equation}
where $\hq_t$ is a $t$-forward measure.\footnote{%
	This is possible under mild assumptions: e.g.~that Fubini's theorem applies, for which $\bb{E}^\hq_0[|e^{-\int_0^t R_s ds}Z_t|]<\infty$ is sufficient, and that $\frac{d\hq_t}{d\hq}:=e^{-\int_0^t ( R_s -\hat R^*_0(s))ds}$ defines a Radon-Nikodym derivative.%
} Now since $\hat R^*_0$ is often observable, the problem of estimating $A_0$ is reduced to estimating this term-structure $\hat Z^*_0$. This is greatly eased by $\hat Z^*_0(0)$ being known and having $\hat Z^*_0(t)\to 0$ as $t\to T$ in practice.\footnote{This can be rationalised as follows. Recalling $Z = (\hl - \ll)U$, we see $Z(T,\cdot)=0$ since $U(T,\cdot)=0$. The continuity of $Z$ and $X_t$ then gives $Z_t\to 0$ a.s.~under $\hq$. Assuming $R_t$ is appropriately bounded from below ($R=R(t,x)$ is typically assumed so in the Feynman-Kac theorem), then $e^{-\int_0^t R_sds}Z_t\to 0$ a.s.~under $\hq$. Adopting conditions of e.g.~dominated convergence gives $\bb{E}_0^\hq\bb[e^{-\int_0^t R_sds}Z_t]\to 0$, and so also $\hat Z_0^*(t):= \bb{E}_0^{\hq_t}\left[Z_t\right]=e^{\int_0^t \hat R_0(s)ds}\bb{E}_0^\hq\bb[e^{-\int_0^t R_sds}Z_t]\to 0$.}

In Section \ref{sec:practical-application} we will approximate this P\&L bleed term-structure by its value under the base measure in order to obtain a tractable expression. So analogous to \eqref{eq:pnl-bleed-defn} we define the quantities
\begin{equation}\label{eq:pnl-bleed-defn-base}
	R^*_0(t) := -\partial_t\log \bb{E}^\qq_0\left[e^{-\int_0^t R_sds}\right],\quad Z^*_0(t) := \bb{E}_0^{\qq_t}\left[Z_t\right].
\end{equation}

\subsection{XVA term-structure}\label{sec:xva-ts}
There remains an apparent difficulty in evaluating $Z^*_0(t)$. If we expand the definition of $Z_t$ from \eqref{eq:fundamental-meta}, we have
$$
Z_t := (\hl - \ll)U_t = (\hat \mu - \mu)\cdot\partial_xU_t + \frac12 (\hat a - a) : \partial_{xx}U_t. 
$$

We wish to avoid having to simulate these pathwise XVA greeks. In analogy to the P\&L bleed term-structure $Z^*_0(t)$ above, we can therefore define an \emph{XVA term-structure} $U^*_0(t):=\bb{E}_0^{\qq_t}[U_t]$. Using the tower property we obtain the computationally convenient representation
\begin{equation}\label{eq:xva-ts-tower}
	U^*_0(t) := \bb{E}^{\qq_t}_0[U_t] = e^{\int_0^t R^*_0(s)ds}\ \bb{E}_0^\qq\left[\int_t^T e^{-\int_0^u R_s ds}F_u du \right] ,
\end{equation}
which essentially comes for free when computing the base XVA $U_0 = U^*_0(0)$ from \eqref{eq:meta-prices}. (It can be obtained simply by truncating the same integral.)

We will ultimately draw an approximate connection between the P\&L bleed term-structure (expectations of weighted future XVA greeks) and greeks of this XVA term-structure.


\subsection{Simple example}

Now we give an essentially minimal extension of Gatheral's adjustment, summarised in \eqref{eq:example-intro}, to a meta-adjustment. To this end, rather than consider Black-Scholes volatility risk on a call option with price $V$, we consider instead Ho-Lee \emph{hazard rate volatility risk} on the option's CVA $U$.\footnote{This Gaussian hazard rate model, allowing $\lambda_t<0$, is clearly unrealistic, but is ideal for this example. In particular, it facilitates a straightforward calibration of $\lambda_t$ to a simple survival curve $P_0(T)=e^{-\lambda_0 T}$ under both $\qq$ and $\hq$. I.e.~$\bb{E}^\hq_0[e^{-\int_0^T \lambda_ tdt}]=\bb{E}^\qq_0[e^{-\int_0^T \lambda_t dt}] =P_0(T)$.} This meta-adjustment can be seen as a way to measure the impact of wrong-way risk on the CVA.\footnote{%
	The approach we follow here can be contrasted with an alternative meta-adjustment approach proposed by \cite{kenyon_2022_wwr}.
}

Denote hazard rate and stock price variables $x=(\lambda,S)$, processes $X_t=(\lambda_t,S_t)$ and functions $V=V(t,S)$, $U=U(t,\lambda,S)$ etc. Let the adjustment $U$ in \eqref{eq:meta-prices} be a simplified CVA:
\begin{equation*}
	U=-\bb{E}_{t,x}^\qq\left[\int_t^Te^{-\int_t^u \lambda_s ds}\lambda_u V^+_u du\right]\quad \text{i.e.}\quad R=\lambda,\quad F=-\lambda V^+,
\end{equation*}
noting that $V^+ = V$ since we are considering a long call option. As in \eqref{eq:meta-prices}, our target CVA $\hat U$ looks identical to $U$, except it is evaluated under $\hq$. Recalling that e.g.~$dX_t=\mu_tdt+\sigma_tdW_t$, we encode (Ho-Lee and Black-Scholes) dynamics of $\lambda_t$ and $S_t$ under $\qq$ and $\hq$ through the functions
\begin{equation*}
	\mu =
	\begin{pmatrix}
		0\\
		0
	\end{pmatrix},\quad
	\sigma =
	\begin{pmatrix}
		0 & 0\\
		0 & \sigma_S S
	\end{pmatrix},\quad
	\hat\mu =
	\begin{pmatrix}
		\hat \sigma_\lambda^2 t\\
		0
	\end{pmatrix},\quad
	\hat\sigma=
	\begin{pmatrix}
		\hat \sigma_\lambda & 0\\
		0 & \sigma_S S
	\end{pmatrix},
\end{equation*}
for constants $\sigma_\lambda,\hat\sigma_\lambda,\sigma_S>0$, along with correlations $\hat\rho=\rho$. Notice that since $dS_t=\sigma_S S_t\,dW^S_t$ under both $\qq$ and $\hq$, our meta-adjustment $A=\hat U-U$ purely captures the impact from hazard rate volatility risk, summarised by the mapping
$$
d\lambda_t=0 \quad \xmapsto{\qq\mapsto\hq}\quad d\lambda_t=\hat\sigma_\lambda^2 t \,dt + \hat\sigma_\lambda dW^\lambda_t.
$$
Putting this into the meta-adjustment representation in \eqref{eq:3}, we have
\begin{equation}\label{eq:meta-adjustment-full}
	A_0 = \bb{E}_0^\hq\left[\int_0^Te^{-\int_0^t \lambda_s ds}Z_t dt\right],
\end{equation}
with P\&L bleed given by
$$
	Z_t = \hat\sigma_\lambda^2 t(\partial_\lambda U)_t + \frac12 \hat\sigma_\lambda^2 (\partial_{\lambda\lambda}U)_t + \rho_{\lambda S}\hat \sigma_\lambda \sigma_S S_t(\partial_{\lambda S}U)_t.
$$

$U$ itself is given by the closed form
$$
U = -\left(1-e^{-\lambda(T-t)}\right)V,\quad V = C\left(S,K,\sigma_S^2(T-t)\right),
$$
for Black-Scholes call price $C$,\footnote{Recall $C(S,K,v) = S N(d_+) - K N(d_-)$ for normal CDF $N$ and $d_\pm=\frac{\log(S/K)}{\sqrt{v}} \pm \frac{\sqrt{v}}{2}$.} and greeks follow by differentiation. Since $\lambda$ is essentially a parameter under $\qq$, with `process' $d\lambda_t=0$, this suggests that meta-adjustments could be used to signal XVA \emph{parameter} risk more generally -- that is, used to estimate the impact of volatility being introduced into XVA model parameters (e.g.~reversion speeds, basis spreads, other volatilities).


\begin{figure}[htbp]
	\centering
	\includegraphics[width=0.7\linewidth]{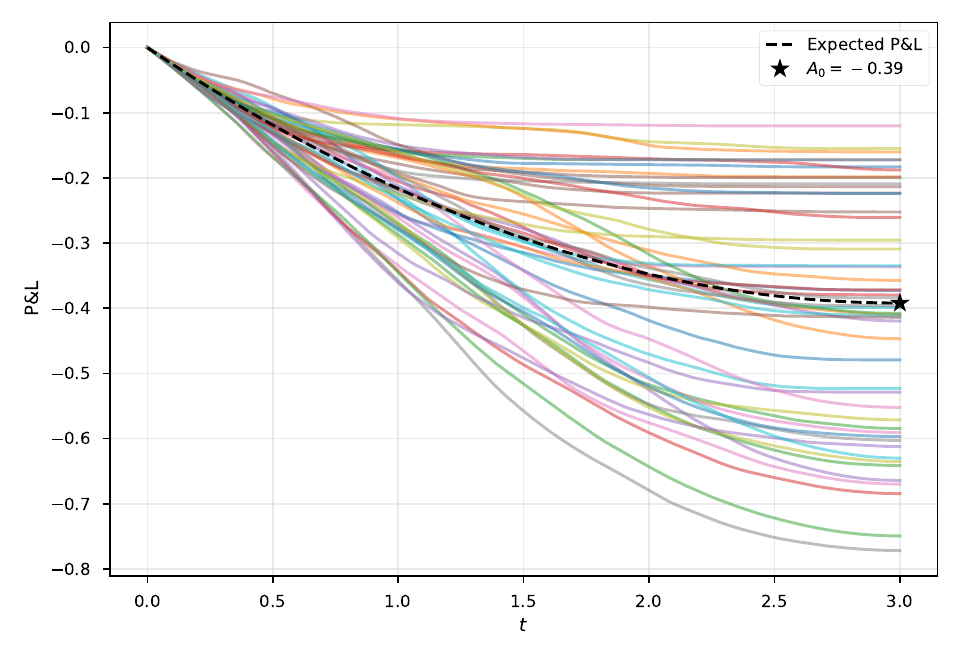}
	\caption{Summarising \eqref{eq:meta-adjustment-full} as $A_0=\bb{E}[\rr{P\&L}_T]$, here we show paths of P\&L as in \cite{ahmad_2005} Figure 3. The expectation of these paths leads to the CVA meta-adjustment $A_0 =-0.393$ as $t\to T=3$.}
	\label{fig:plot}
\end{figure}

This example can be validated succinctly, without requiring advanced numerical methods for simulating future P\&L bleeds via CVA greeks. As in Figure \ref{fig:plot}, this leads to the meta-adjustment $A_0$ via \eqref{eq:meta-adjustment-full}. With inputs $\rho_{\lambda S}=0.9$, $\hat\sigma_\lambda=0.01$, $\sigma_S=0.2$, $\lambda_0=0.05$, $S_0=100$, $K=100$ and $T=3$, we obtain call price $V_0=13.75$, base CVA $U_0=-1.92$ and CVA meta-adjustment $A_0=-0.393$, using 100,000 paths and 1,000 time steps. Evaluating the target CVA directly confirms $\hat U_0=U_0 + A_0$, although doing so via the meta-adjustment $A_0$ provides a significant variance reduction.

\subsection{Practical application}\label{sec:practical-application}
The above example could be solved analytically, but in general (a) we do not have an analytic form for $U$ that lends itself to a perfect expression of $Z$ in future market states, and (b) we cannot run a simulation in the target measure $\hq$. In that case we must make approximations; this is where the term-structure concepts introduced above begin to show their value.

We will make two approximations, which together give us a version of \eqref{eq:fundamental-meta} that can be calculated with any framework that can compute the base price $U$:
\begin{enumerate}
	\item We approximate measure $\hq$ with $\qq$ for calculating the meta-adjustment itself: $\bb{E}_{t,x}^\hq\left[\int_t^T e^{-\int_t^u R_sds}\bleed_udu \right] \approx \bb{E}_{t,x}^\qq\left[\int_t^T e^{-\int_t^u R_sds}\bleed_udu \right]$.
	\item We approximate simulated operators by their day-zero values, via term-structures: $Z^*_0(t) := \bb{E}^{\qq_t}_0\left[(\hl - \ll) U_t\right]\approx (\hl - \ll) \bb{E}^{\qq_t}_0[ U_t] =: (\hl - \ll)U^*_0(t)$. (Operators applied to $U^*_0(t)$ here should be read as `shocking' the day-zero market -- e.g.\ $\partial_x U^*_0(t) \equiv \partial_x \bb{E}_{0,x}^{\qq_t}[U_t]|_{x=X_0}$.)
\end{enumerate}
Together, the above steps enable us to simplify
\begin{equation}
	A_0 \approx \bb{E}_0^\qq\left[\int_0^Te^{-\int_0^t R_s ds} Z_t \,dt\right] 
	= \int_0^T e^{-\int_0^t R^*_0(s)ds} Z^*_0(t) \,dt , 
\end{equation}
finally giving us a fully tractable and general approximation of the meta-adjustment:
\setlength{\fboxsep}{2mm}
\begin{equation}\label{eq:computable}
	\boxed{
	A_0 \approx \int_0^T e^{-\int_0^t R^*_0(s)ds} (\hl - \ll)U^*_0(t) \, dt . 
	} 
\end{equation}
Applying this very general expression to the wrong-way risk problem of \eqref{eq:meta-adjustment-full} we have the very computable expression
\begin{equation}\label{eq:meta-adjustment-approx}
	A_0 \approx \int_0^T e^{-\lambda_0 t} \left(
	\hat\sigma_\lambda^2 t\, \frac{\pd}{\pd \lambda_0} 
	+ \frac12 \hat\sigma_\lambda^2 \frac{\pd^2}{\pd \lambda_0^2} 
	+ \rho_{\lambda S}\hat \sigma_\lambda \sigma_S S_0 \frac{\pd^2}{\pd\lambda_0 \pd S_0} \right) U^*_0(t) \,dt.
\end{equation}

This case can actually be expressed in a suggestive closed form, as we will do below.\footnote{%
	A step towards this is recognising that in this case we have simply $U^*_0(t) = -\left(1-e^{-\lambda_0(T-t)}\right)V_0$.%
} However, that must not distract from the key point: \eqref{eq:computable} in general provides an estimate for the pricing adjustment that can be calculated by any framework that supports the calculation of the base value. The above notwithstanding, an analytic form of \eqref{eq:meta-adjustment-approx} is:\footnote{%
	In this case the approximation matches that which would be obtained via a first-order Taylor expansion on the credit volatility. However, unlike the method we present, a Taylor-based approach is generally impractical, as one cannot normally calculate sensitivities with respect to a volatility that is not present in the base calculation.%
}
\begin{equation}
	A_0 \approx -\frac{1}{2} \rho_{\lambda S} \hat\sigma_\lambda \sigma_S S_0 T^2 e^{- \lambda_0 T} \frac{\partial V_0}{\partial S_0}.
\end{equation}
For the same parameter values as before, this gives an estimate of $-0.397$, representing an error of just 1\% on the true value of $A_0 = -0.393$ and demonstrating the power of the approach presented here.

\begin{figure}[htbp]
	\centering
	\includegraphics[width=0.7\linewidth]{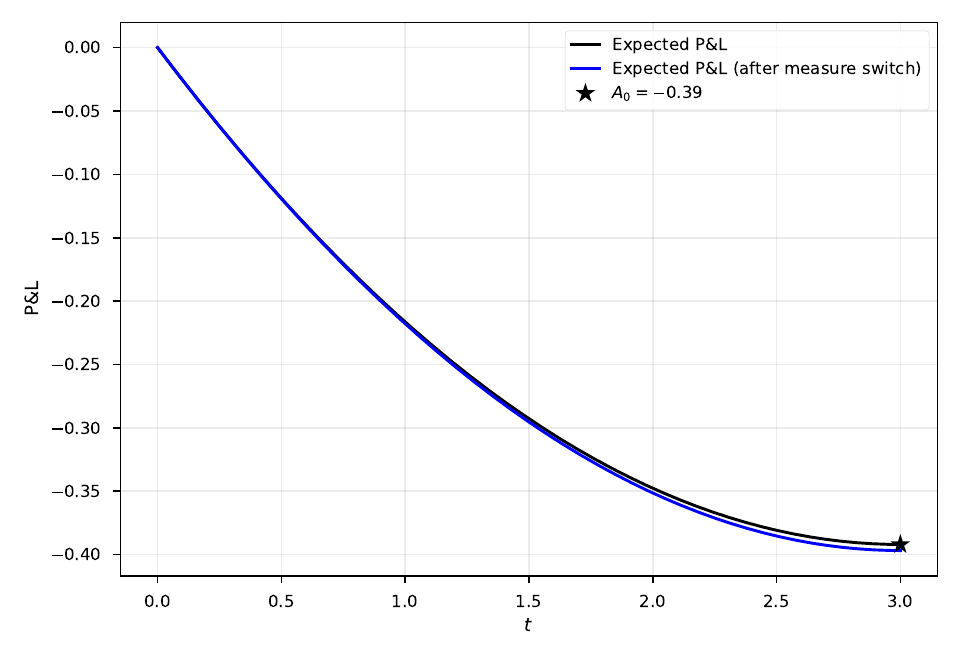}
	\caption{A repeat of the expected path from Figure \ref{fig:plot}, corresponding to the expected P\&L implied by exact calculation, along with the effect of our approximations.}
	\label{fig:approximations}
\end{figure}

To provide some intuition regarding the impacts of the approximations involved, Figure \ref{fig:approximations} shows the progression of the expected P\&L (as seen in Figure \ref{fig:plot}) via precise calculation, and after the approximations are applied. In this example case the approximation via day-zero quantities has no impact, and so all the inaccuracy stems from the approximation of the measure. The plot makes it clear just how good that approximation is in the present case.

\section{Conclusion}\label{sec:conclusion}

In \eqref{eq:fundamental-representation}, this article provides a \emph{fundamental representation} of pricing adjustments, summarised in \eqref{eq:intro-adjustment}. We have shown in Section \ref{sec:examples} how this representation encapsulates a rich history of pricing adjustments, including XVA. In Section \ref{sec:meta} we showed how this can be put to work on new challenges such as handling XVA model risk, culminating in a very practical approximation in \eqref{eq:computable}. 

Importantly, the representation is not unduly complicated. It essentially says that all pricing adjustments take the form of an expected discounted future P\&L bleed, and that this bleed breaks down into \emph{model}, \emph{discounting} and \emph{payoff} parts. These ingredients are \emph{natural} given that they define our pricing functions in the first place; they reveal the simple classification of pricing adjustments depicted in Figure \ref{fig:venn} and have relatively clear historical exemplars (Section \ref{sec:examples}).

The topic of meta-adjustments from Section \ref{sec:meta} warrants deeper numerical analysis, with applications to XVA model/parameter risk especially. This is a hot topic in both academia and industry, given the work of e.g.~\cite{silotto_2024} and \cite{benezet_2024}, and increased regulatory awareness and scrutiny. The present article provides a robust framework for tackling this issue, bringing rigour and accuracy to the type of adjustments or reserves that good traders should already consider in the light of possible future P\&L.

\vspace{2mm}
\subsection*{Acknowledgements}\label{sec:acknowledge}

The authors have benefited from many discussions over several years with members of Barclays's XVA Quantitative Analytics (QA), Model Validation and Trading teams, especially Raphael Albrecht, Simon O'Callaghan, Hamza Saber, Alex White and Ieuan Williams. Related material was presented and discussed at a QA Open Forum in November 2023, and at QA XVA team talks.

\clearpage
\phantomsection
\addcontentsline{toc}{section}{References}
\bibliographystyle{apa}
\renewcommand\refname{References}
\bibliography{references}

\clearpage
\appendix
\section{Pathwise P\&L}\label{app:pnl}

Recall the fundamental representation of adjustments as in \eqref{eq:fundamental}, i.e.
\begin{equation}\label{eq:fundamental-app}
	U=\bb{E}_{t,x}^\hq\left[\int_t^Te^{-\int_t^u\hat R_s ds}\bleed_u du\right],\quad \bleed := (\hl - \ll)V - (\hat R - R)V + (\hat F - F).
\end{equation}
The main purpose of this appendix is to clarify why the function $\bleed$ should be interpreted as a hedged portfolio's P\&L bleed, and therefore why the variable $\int_t^Te^{-\int_t^u\hat R_s ds}\bleed_u du$ above should be understood as a pathwise P\&L. As discussed in Section \ref{sec:intro}, this is clearly a natural candidate on which to base risk signals, in the absence of being able to compute $U$ efficiently and accurately, using $\hq$.

Towards this, we should not simply adopt the target price $\hat V$ and its dependencies $\hl$, $\hat R$ etc.~like we do elsewhere. We should rather provide interpretations of these. The most difficult to explain is the `risk-neutral' drift $\hat\mu$. Notice that this appears twice in \eqref{eq:fundamental-app}: both in $\bleed$ via $\hl$, which can be legitimately evaluated under \emph{any} $\hp\neq\hq$, and in the drift of $X_t$, \emph{specifically} under $\hq$.

There is however no \emph{canonical} P\&L interpretation, because this depends on the \emph{assumed} collateral, hedging, funding etc.\footnote{This ambiguity originates from the many-to-one nature of the map from collateral, hedging, funding etc.~to prices. We trust that readers can adapt the presentation here as required.} Surprisingly, we are able to follow the classical setup of \cite{ellersgaard_2017}, e.g.~\emph{without} default or collateral, nor rehypothecation.\footnote{Still, effects of e.g.~\emph{hedged} default and collateral may be considered to reside within $\hat F$, likewise transaction costs etc. On the other hand, hedge rehypothecation influences $\hat\mu$.}  Two generalisation are necessary, however: (a)~$V$ takes the form in \eqref{eq:base-price-sde}, so can represent an \emph{exotic} derivative, even an XVA; and (b)~the risk factors in $X_t$ are not assumed to be tradable.

\subsection*{Details}

Now consider the cumulative P\&L $\Pi$ of a hedged portfolio, in light of a measure $\hp$ which reflects the \emph{real} evolution of the process $X_t$. This evolution is given by
$$
d X_t=\hat \mu^*dt + \hat \sigma d W_t,\quad d\langle W\rangle_t=\hat\rho dt
$$
where we e.g.~abbreviate $\hat\sigma=\hat\sigma_t=\hat\sigma(t,X_t)$ here. Notice the real drift $\hat \mu^*\neq\hat\mu$. For weights $\Delta$ of hedge instruments $H$, we then have portfolio value $V+\Delta H+B=0$,\footnote{I.e.~$B := -V-\Delta H$. Notice $\Pi$ is \emph{not} the portfolio value $V+\Delta H+B$. In the terminology of \cite{duffie_2001} Section 6.L, $\Pi$ is the cumulative \emph{dividend process generated by $\Delta$}. Here $\Delta$ is a row vector and $H$ a column vector of prices; we assume that such things can be inferred.} for funding account $B$ with rate $\hat R$. If we can choose $\Delta$ to hedge the diffusive component of $dV$, and thereby make the incremental P\&L $d\Pi$ of order $dt$ at most (in the absence of unhedged jumps), then we eventually obtain
\begin{equation}\label{eq:incremental-pnl}
	d\Pi = dV - \hat R V dt +\hat Fdt -\partial_x V(d X_t - \hat\mu dt).
\end{equation}

This expression coincides with \cite{ellersgaard_2017} Equation (6), except that $V$ yields the additional payoff $\hat F$, and the drift $\hat \mu$ is complicated by $X_t$ not being tradable. In general, towards \eqref{eq:incremental-pnl} we should use the framework of \cite{duffie_2001} Section 6.L, treating hedges $H$ like $V$, and first arriving at
\begin{equation}\label{eq:hedged-pathwise-pnl}
	d\Pi = dV-\hat RVdt	+\hat Fdt+\Delta (dH - \hat R Hdt + \hat F^*dt)
\end{equation}
for hedge payoffs $\hat F^*$.\footnote{Practitioners are familiar with such P\&L expressions, but they are not trivial to derive. This expression for $d\Pi$ is an incremental version of $D^\theta_t=\theta_0\cdot X_0 + \int_0^t \theta_s dG_s - \theta_t \cdot X_t$ in \cite{duffie_2001} Section 6.L, where $D^\theta$ is the \emph{cumulative dividend process generated by} $\theta$, and for us $\theta_t \cdot X_t = \theta_0 \cdot X_0 = 0$. We do not see terms like $d\theta_s$ here because these are collectively zero (by so-called self-financing). We recommend following \cite{shreve_2004} Exercise 4.10 as a starting point. In particular, the balance $B$ should be treated as an investment weight $\Gamma$ in a money market price process $M$, that grows according to $dM=\hat RMdt$. Then $d\Pi = (dV+\hat Fdt) + \Delta(dH + \hat F^*dt) + \Gamma (dM)$, and \eqref{eq:hedged-pathwise-pnl} follows from $\Gamma M = B = -V-\Delta H$.} Now setting $\Delta = -\partial_x V (\partial_x H)^{-1}$ and applying It\^o's lemma to $dH$, we obtain \eqref{eq:incremental-pnl} specifically with the peculiar drift
\begin{equation*}
	\hat\mu = (\partial_x H)^{-1}\left(\hat R H - \hat F^* - \partial_t H - \frac12\partial_{xx} H:\hat a\right).\footnote{Here $(\partial_x H)_{ij}=\partial_j H_i$ is the Jacobian matrix and $(\partial_{xx} H)_{ijk}=\partial_{jk}H_i$ the Hessian \emph{tensor}.}
\end{equation*}
This reduces to $\hat\mu = \hat R X_t - \hat F^*$, consistent with \cite{ellersgaard_2017}, when $X_t$ is tradable and specifically $H=X_t$, since then $\partial_x H=1$ and $\partial_t H=\partial_{xx} H=0$. Otherwise, this expression for $\hat\mu$ appears quite unsatisfactory. However, rearranging it into a familiar parabolic PDE (actually PDEs) for $H$, i.e.
\begin{equation*}
	\partial_t H + \hat{\cc{L}} H -\hat RH	+\hat F^* =0,\quad \hat{\cc{L}}:=\hat\mu\cdot\partial_x + \frac12 \hat a:\partial_{xx},
\end{equation*}
exposes the interpretation of $\hat\mu$: it is the drift of $X_t$ under a measure $\hq\sim\hp$ where the (observable) hedge prices $H$ admit the usual stochastic representations as expected discounted payoffs.\footnote{As in \cite{piterbarg_2010}, if (full) hedge rehypothecation is possible at repo rates $\hat R^*$, then $\hat R H$ everywhere becomes $\hat R^* H$. So in particular hedges instead drift at $\hat R^* H-\hat F^*$ under $\hq$.} Under $\hq$, hedges themselves drift at $\hat R H - \hat F^*$,
\begin{equation*}
	dH = \partial_t Hdt + \partial_x H dX_t + \frac12 \partial_{xx} H :d\langle X\rangle_t	= \underbrace{(\partial_t + \hat{\cc{L}})H}_{\hat R H - \hat F^*}dt + \partial_x H\hat\sigma d W_t.
\end{equation*}

Despite this interpretation, note that \eqref{eq:incremental-pnl} remains under $\hp$, e.g.~$\hat\mu$ is evaluated using $X_t$ with \emph{real} drift $\hat\mu^*\neq\hat\mu$. Returning to \eqref{eq:incremental-pnl}, write
\begin{equation*}
	dV = \partial_t Vdt + \partial_x V dX_t + \frac12 \partial_{xx} V :d\langle X\rangle_t
	= \left(\partial_t + \frac12 \hat a:\partial_{xx}\right)V dt + \partial_x V dX_t.
\end{equation*}
Then after eliminating the theta $\partial_t V$ via the PDE in \eqref{eq:base-pde}, the apparent incremental P\&L takes on the familiar form in terms of our P\&L bleed,\footnote{Since $d\Pi_t=\bleed_t dt$, then $\bleed_t=\frac{d\Pi}{dt}$ a.s. This explains our (P\&L \emph{bleed}) description of $\bleed$. }
\begin{equation*}
	d\Pi = (\hl - \ll)Vdt - (\hat R - R)Vdt + (\hat F - F)dt =:\bleed dt.
\end{equation*}

Now we view this as defining the payoff of an exotic derivative, and standard pricing arguments, via another hedge portfolio, then tells us that zero net P\&L will be achieved if $U$ satisfies the PDE in \eqref{eq:adjustment-pde}. This leads us to the stochastic (fundamental) representation for the adjustment $U$ in \eqref{eq:fundamental-app} under $\hq$, now via the P\&L bleed $\bleed$ experienced without such an adjustment.

\clearpage
\section{Model dependence}\label{app:model-dependence}

We have worked throughout without exposing parameters, e.g.~by writing $V^\theta=V(t,x;\theta)$, $V^\theta_t=V(t,X_t;\theta)$, etc. It can be especially helpful to expose \emph{model} parameters $\theta\in\bb{R}^m$ dictating processes $X_t$ under measures $\qq^\theta$ according to
\begin{equation*}
	d X_t = \mu_t^\theta dt + \sigma_t^\theta d W_t.
\end{equation*}
Indeed we can now w.l.o.g.~consider model differences via parameters $\hat\theta\neq\theta$, in place of the previous function e.g.~$\hat\sigma\neq\sigma$ and thus generator $\hl\neq\ll$ differences.

Now we can explicitly expose additional model dependence in $V^\theta$ coming purely from a dependence of $R^\theta$, $F^\theta$ etc.~on $\theta$, so e.g.~\eqref{eq:base-price-sde} becomes
\begin{equation*}
	V^\theta = \bb{E}_{t,x}^{\qq^\theta}\left[\int_t^T e^{-\int_t^u R^\theta_s ds}F^\theta_u du + e^{-\int_t^T R^\theta_u du} G^\theta_T  \right].
\end{equation*}
This leads to a defensive alternative exposition of pricing adjustments, which exposes model dependence in functions like $F$, or more specifically \emph{in differences} e.g.~$\hat F-F$ appearing in \eqref{eq:intro-adjustment}, which might be overlooked otherwise.

To be clear, this alternative exposition is equivalent mathematically, but debatably provides a more refined separation of the pricing adjustments universe in Figure \ref{fig:venn}, where model adjustments $\hl\neq\ll$ are \emph{equivalently} replaced by those where $\hat\theta\neq\theta$, but e.g.~$\hat F\neq F$ means $\hat F(t,x;\theta)\neq F(t,x;\theta)$ more specifically. This doesn't change the collection of adjustments in e.g.~M or $\rr{P}-\rr{M}$, but the intersection $\rr{M}\cap\rr{P}$ shrinks to exclude cases where $\hat F(t,x;\theta)= F(t,x;\theta)$.

The price of this refinement is that we lose the elegant correspondence emerging from \eqref{eq:intro-adjustment}, between the sets M, D and P in Figure \ref{fig:venn}, and the differences $\hl - \ll$, $\hat R - R$ and $\hat F - F$ which contribute to the P\&L bleed $\bleed$.

\subsection*{Exercise}

Gatheral's example from Section \ref{sec:gatheral} can be modified slightly to expose model dependence. First notice that the use of functions $\hat G=G$ quietly limited the scope of this example to model-\emph{independent} terminal payoffs. It is instructive to consider a (model dependent) terminal payoff at \emph{earlier} time $\tau<T$, where the payoff is given by the \emph{price} of the same European option with maturity $T$. For the Black-Scholes call option case, the terminal payoffs at time $\tau$ become
\begin{multline*}
	G(S)= \bb{E}^\qq_{\tau,S}[(S_T-K)_+]=C(S,K;\alpha^2(T-\tau))\neq\\ C(S,K;\hat\alpha^2(T-\tau))=\bb{E}_{\tau,S}^\hq[(S_T-K)_+]=\hat G(S).
\end{multline*}
Now it is clear that $\hat G=\hat G(S;\hat\alpha)\neq G(S;\alpha)= G$, i.e.~$G^*:=\hat G - G\neq0$, even though $\hat G(S;\alpha)= G(S;\alpha)$. We leave it as an exercise to write out the resulting fundamental adjustment as in \eqref{eq:succinct-repr} in full, which takes the form

\begin{equation*}
	U = \bb{E}_{t,x}^\hq\left[\int_t^\tau (\hl-\ll)V_udu + (\hat G - G)_\tau\right],
\end{equation*}
and is actually invariant to the time $\tau\in[t,T]$. This coincides with Gatheral's adjustment in \eqref{eq:local-vol-adj} when $\tau=T$, and so $\hat G=G$ is model-\emph{independent}.

In practice such model dependent \emph{terminal} payoffs are the exception. However, counterpart payoffs $\hat F=\hat F(t,x;\hat \theta)\neq F(t,x;\theta)=F$ are the norm in XVA, because these typically depend on the value of derivatives at such times $\tau<T$.

\end{document}

%% file: picture.tex

\def\exa{\cite{gatheral_2006}}
\def\exb{\cite{piterbarg_2010}}
\def\exc{\cite{burgard_2013}}

\def\disttocirc{1.5}
\def\baseangle{15}
\def\anglea{\baseangle+90}
\def\angleb{\anglea+120}
\def\anglec{\angleb+120}
\def\radius{2}
\def\textshift{.5}
\def\coloura{purple}
\def\colourb{teal}
\def\colourc{blue}
	\begin{tikzpicture}
	\node [shift={(\anglea:\disttocirc)}] (centrea) {};
	\node [shift={(\angleb:\disttocirc)}] (centreb) {};
	\node [shift={(\anglec:\disttocirc)}] (centrec) {};
	\begin{scope}[line width=0.5mm, opacity=.4, fill opacity=.05]
		\draw[draw=\coloura, fill=\coloura] (centrea) circle (\radius);
		\draw[draw=\colourb, fill=\colourb] (centreb) circle (\radius);
		\draw[draw=\colourc, fill=\colourc] (centrec) circle (\radius);
	\end{scope}
	\begin{scope}[text width=2cm, align=center]
		\node[shift={(\anglea:\textshift-.3)}] at (centrea) {\textbf{M}odel adjustments $\hl \neq \ll$};
		\node[shift={(\angleb:\textshift)}]    at (centreb) {\textbf{D}iscounting adjustments $\hat R \neq R$};
		\node[shift={(\anglec:\textshift-.05)}]    at (centrec) {\textbf{P}ayoff adjustments $\hat F \neq F$};
	\end{scope}
	\begin{scope}[opacity=.5]
		\node[shift={( 20:1.7)}] (targeta) at (centrea) {$\circ$};
		\node[shift={(150:1.7)}] (targetb) at (centreb) {$\circ$};
		\node[shift={(235:1.7)}] (targetc) at (centrec) {$\circ$};
		
		\node[] (citea) at (2.7,2.5) {\exa};
		\node[] (citeb) at (-3.8,0.7) {\exb};
		\node[] (citec) at (2.2,-2.9) {\exc};
		
	\end{scope}
	\begin{scope}[opacity=.25]		
		\draw[->] (citea) edge[bend left=30] (targeta);
		\draw[->] (citeb) edge[bend right=30] (targetb);
		\draw[->] (citec) edge[bend right=30] (targetc);
	\end{scope}
\end{tikzpicture}